\documentstyle[epsf]{article}
\begin{document}

\thispagestyle{empty}
\parskip=12pt
\raggedbottom

\def\mytoday#1{{ } \ifcase\month \or
 January\or February\or March\or April\or May\or June\or
 July\or August\or September\or October\or November\or December\fi
 \space \number\year}
\noindent
\begin{flushright}
 COLO-HEP-384 \\ KLTE-DTP/1997/3 
\end{flushright}

\begin{center}
{\Large \bf  Remarks on abelian dominance}
\footnote{Work supported in part by U.S.\ Department of Energy grant
DE-FG02-92ER-40672 and OTKA Hungarian Science Foundation T 017311.}

\vspace{1cm}

Tam\'as G.\ Kov\'acs \\
Department of Physics \\
University of Colorado, Boulder CO 80309-390, USA \\[5mm]

Zsolt Schram \\
Department of Theoretical Physics, Kossuth Lajos University \\
Debrecen H-4010, Hungary

\vspace{0.5cm}


\end{center}

\nopagebreak[4]

\section*{Abstract}

We used a renormalisation group based smoothing to address two questions
related to abelian dominance. Smoothing drastically reduces short distance
fluctuations but it preserves the long distance physical properties of the
SU(2) configurations. This enabled us to extract the
abelian heavy-quark potential from time-like Wilson loops 
on Polyakov gauge projected  configurations. We obtained a very 
small string tension which is inconsistent with the string tension 
extracted from Polyakov loop correlators. This shows that the Polyakov 
gauge projected abelian configurations do not have a consistent 
physical meaning. We also applied the smoothing on SU(2) configurations 
to test how sensitive abelian dominance in the maximal abelian gauge 
is to the short distance fluctuations. We found that on smoothed 
SU(2) configurations the abelian string tension was about 30\% smaller 
than the SU(2) string tension which was unaffected by smoothing. This
suggests that the approximate abelian dominance found with the Wilson 
action is probably an accident and it has no fundamental physical 
relevance.

\eject


\section{Introduction}

It is an old idea to try to understand non-abelian gauge theories
in terms of an effective abelian model with a smaller symmetry
group. One possible way of doing this on the lattice is to isolate
$U(1)^{N-1}$ link variables belonging to a maximal torus of SU(N).
This is called abelian projection.
The hope is that non-abelian confinement might be explained as a 
condensation of monopoles in the resulting abelian projected model
(see e.g.\ \cite{Polikarpov} for a recent review).
If one wants to explain the non-abelian physics in the abelian 
projected system, a necessary condition is that the abelian model
has to reproduce the physical features of the non-abelian system.
This property is referred to as abelian dominance.

Since the abelian symmetry group is smaller than the non-abelian
one, this procedure necessarily involves some gauge fixing before
the projection is done. In principle the physical properties of
the projected system can depend on the gauge choice. This was originally
perceived as a nuisance and there are still attempts to prove
that the physical properties of the projected system are independent
of the gauge choice (see e.g.\ \cite{Ejiri}). 
Up to now the only gauge in which abelian 
projected system seems to capture the physics of the non-abelian 
model is the maximal abelian gauge \cite{MAG}.
Here in the SU(2) case the abelian and non-abelian 
string tensions at Wilson $\beta=2.51$ agree to within 8\% \cite{Bornyakov}. 
In other gauges, most notably in the Polyakov gauge (where Polyakov 
loops are diagonalised) the situation is more controversial. 
Since all the Polyakov loops can be exactly diagonalised at the 
same time, in this case ``abelian dominance'' exactly and trivially
holds if the string tension is measured with Polyakov loop correlators.
On the other hand due to the high level of noise on the projected 
configurations, it is impossible to extract the string tension 
from Wilson loops \cite{Suzuki,Schram}. 

In the present paper we study some related issues.
The first question we address is that of the gauge 
choice. We use a recently proposed smoothing technique based on
renormalisation group ideas \cite{DeGrand}. We can drastically 
reduce the short-distance fluctuations while preserving the 
long-distance physical properties of our configurations, most
importantly the SU(2) string tension. This allows us to extract 
the heavy quark potential from Wilson loops on Polyakov gauge projected 
configurations. Doing the gauge fixing and the projection 
on the smoothed configurations, the resulting abelian string
tension turns out to be practically zero. This result is inconsistent
with the string tension measured from Polyakov loop correlators.
It shows that the physical meaning of Polyakov gauge projected 
configurations is questionable. 
We argue on general grounds that probably some other gauge
choices also suffer from the same problem and we also present
a set of minimal requirements that a gauge choice has to satisfy
in order to avoid this problem. We conclude that it is natural 
to expect that the gauge choice has a strong influence on the 
physical properties of the abelian configurations. We suggest that
rather than trying to prove the gauge independence of the projection,
one has to concentrate on finding one particular gauge in which 
abelian dominance holds. The only gauge known to us in which approximate
abelian dominance has been found (with the Wilson action) is the maximal
abelian one. Therefore in the second part of the present paper we shall 
concentrate only on this gauge.

The approximate nature of the agreement between the abelian and 
non-abelian string tension raises some doubts as to whether it is a 
real physical effect or an accident.
In particular, if abelian dominance is a genuine physical effect,
it should hold in the continuum limit and also it should be universal.
We can look at the abelian and non-abelian string 
tensions as two different physical quantities with the same mass
dimension. Their ratio in the continuum limit has to be unique and 
if abelian dominance holds, close to unity. Unfortunately a detailed scaling test 
on any abelian quantity would be extremely expensive in an iteratively
fixed gauge like the maximal abelian one. In the present work we have
a much more modest aim. We study a related question, how abelian
dominance depends on the details of the short-distance fluctuations. 
Using the above mentioned smoothing on Monte Carlo 
generated SU(2) gauge configurations
we can produce smoothed configurations with the same long-distance
properties but reduced short-distance fluctuations. Comparing the
abelian string tension on the original and the smoothed configurations
we can gain insight into its dependence on the short-distance details.
We find that the abelian string tension is very sensitive to
the short distance structure dropping by about 30\% after one smoothing
step. This raises the question of how much of the abelian string
tension comes from genuine SU(2) long distance physics and how much of 
it is a reflection of physically irrelevant short range fluctuations.

The plan of the paper is as follows. In Section 2 we briefly describe
the smoothing procedure that will be used. In Section 3 we address 
the question of the gauge choice, present our results about the string
tension on Polyakov gauge projected configurations and make some
general remarks. In Section 4 we study the
question of universality of abelian dominance in the maximally 
abelian gauge. Finally in Section 5 we present our conclusions.

\section{Smoothing}

In this section we describe the  main idea of the smoothing procedure 
that we use. A more detailed account can be found in Ref.\ \cite{DeGrand}.
Let us consider a scale two real space renormalisation
group transformation (blocking) that maps the original (fine)
lattice on a coarser lattice with twice the lattice spacing 
and $2^4$ (in $d=4$) times less degrees of freedom. By construction
blocking preserves all physical features (correlations) of
the fine lattice on distance scales larger than the coarse lattice
spacing. Since blocking is a coarse graining procedure and there
are many fine configurations which are mapped on the same 
coarse one, it has no inverse. Nevertheless one can define
an opposite operation, we call it inverse blocking, which 
assigns to any coarse configuration the smoothest (smallest
action) of all those fine configurations that block into it.
Inverse blocking can be thought of as interpolating to a finer
lattice in the smoothest possible way while preserving all
the physical features of the coarse configuration. Now we can 
describe the smoothing procedure. 

One step of smoothing consists
of an inverse blocking on a finer lattice followed by a blocking
but on a different coarse sublattice of the fine lattice. Using
a different sublattice is essential because otherwise the 
fine configuration would just block back into the same coarse
configuration that we started with. The crucial point is that if
the original physical lattice spacing was $a$ and the bare coupling
$g(a)$ then the inverse blocked configurations will be locally a lot 
smoother than typical configurations at a coupling $g(a/2)$.
This is because inverse blocking  selects the lowest action 
configuration from among all the ones that would block into 
the given coarse configuration. Now a blocking step on a different
coarse sublattice will treat the fine configuration as if it
corresponded to a coupling larger than $g(a/2)$ and block it 
into a configuration with effective bare coupling larger than 
$g(a)$. This is however not true if the blocking is performed
on the original coarse sublattice since in this case by 
construction we would get back the original configuration. 
This happens because the inverse blocked configuration has a
certain staggered structure that ``remembers'' the fact that 
it came form a coarse lattice by inverse blocking, this is 
why a certain sublattice is distinguished.

The net result of one 
smoothing step is that the lattice returns to its original
size, it has essentially unchanged long-distance physical content
(since both blocking and inverse blocking preserve this) but the
shortest distance fluctuations and consequently the action 
are considerably reduced. In Ref.\ \cite{DeGrand} after a few
such smoothing steps the action was seen to drop by almost 
two orders of magnitude while the string tension and the 
instanton content remained the same. On the other hand, the
additive constant term of the heavy quark potential extracted 
from Wilson loops decreased considerably, reflecting the fact
that much of the short distance fluctuations have been 
removed.

Performing several successive smoothing steps will gradually 
reduce fluctuations at larger and larger distance scales but
it does not affect the genuine asymptotic long-distance 
observables. In Table \ref{tab:cycling} we illustrate how effective
this smoothing procedure is in removing short-distance fluctuations.
In this Table we present how the average plaquette, the action
and the string tension changes with the smoothing. The measurements
were done on the same ensemble of 100 $8^3 \times 12$ 
fixed point action $\beta=1.5$ configurations that we also use
in Section 4.

\begin{table*}[hbt]
\setlength{\tabcolsep}{1.5pc}
\caption{The average plaquette, the action and the string tension
after smoothing. Step 0 refers to the original $8^3 \times 12$ 
configurations generated with a fixed point action at $\beta=1.5$.}
\label{tab:cycling}
\begin{tabular*}{\textwidth}{@{}l@{\extracolsep{\fill}}lcccccc}
\hline
smoothing step   & 0  & 1 & 2 & 3  \\
\hline
 plaquette  &  1.030 &  1.908  & 1.960  & 1.972 \\
 action   &  35000 & 3100  & 1400 &  960    \\
 $\sigma a^2$  &  0.123(7)  &  0.115(9) & 0.112(7)  &  0.118(6)  \\
\hline
\end{tabular*}
\end{table*}

The blocking kernel that we used is completely 
identical to the one of Ref.\ \cite{DeGrand}. All of our
configurations were generated with the fixed point action of
Ref.\ \cite{DeGrand} because this is the fixed point action 
corresponding to our RG transformation and therefore it is 
consistent with the blocking and inverse blocking.

\section{The gauge choice}

The very idea of abelian 
dominance is that the diagonal abelian degrees of freedom can account
for the physical properties of the full non-abelian configurations.
The issue of gauge fixing is definitely important here since the
part of the system that we retain/discard with the abelian projection
very strongly depends on it. There are several gauge fixing proposals
in the literature in this context. The idea behind some of them is 
that a certain class of operators can be made diagonal by a suitable
gauge fixing and for these operators the results obtained after abelian 
projection are exactly equal to the full non-abelian measurements.
This is taken as evidence for abelian dominance (see e.g.\ \cite{Ejiri}
for the case of Polyakov loops in the Polyakov gauge). 

We want to emphasise that this is only
a trivial consequence of the non-abelian gauge freedom and it does not
mean that the abelian part reproduces the relevant physical properties
of the full system. On any given SU(2) configuration all
the links belonging to the Polyakov loops can be diagonalised 
simultaneously by a suitable gauge transformation. Therefore any 
physical quantity derived from Polyakov loops will be trivially 
and exactly reproduced after abelian projection in this gauge. 
In particular there is exact abelian dominance for the string 
tension measured with Polyakov loop correlators \cite{Ejiri}.

A good test of whether the Polyakov gauge projected abelian 
configurations capture some genuine physics would be to 
measure the string tension using time-like Wilson loops and
compare this to the string tension obtained with Polyakov loop
correlators. Unfortunately this cannot be done directly because
the gauge fixing introduces so much noise that one would need
a huge number of configurations to get enough statistics. 

We can however use an ensemble of smoothed configurations
and do all the measurements on them. Since smoothing does not
change anything that is genuine long distance physics, this is
perfectly justified. In fact we could look at the smoothed 
configurations as another ensemble generated with some unknown action
which produces different short distance structure but its 
$\beta$ value is adjusted to be at the same physical lattice
spacing (fixed e.g. with the string tension) as the 
unsmoothed configurations.

\begin{figure}[!htb]
\begin{center}
\vskip 10mm
\leavevmode
\epsfxsize=110mm
\epsfbox{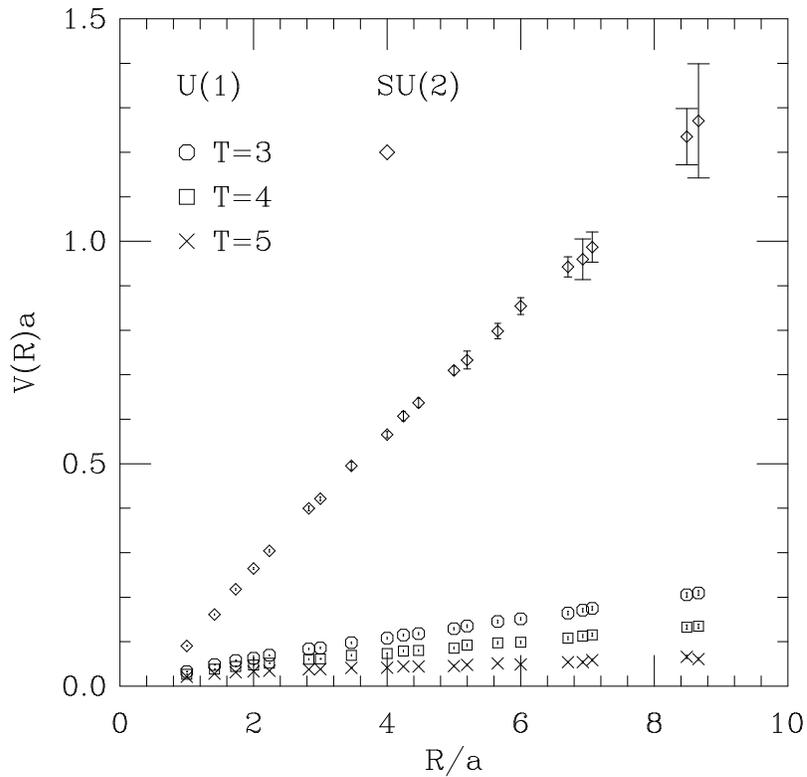}
\end{center}
\caption{The static quark potential measured with timelike Wilson
loops. Diamonds correspond to the full SU(2) potential, the other 
three symbols represent the U(1) potential measured in the Polyakov 
gauge with Wilson loops of different time extensions.}
\label{fig:pot_pg}
\end{figure}

We generated an ensemble of 20 $12^4$ configurations with the
fixed point action of Ref.\ \cite{DeGrand} at $\beta=1.5$ which
corresponds to a physical lattice spacing of 0.144 fm. After one
smoothing step we measured both the full SU(2) and 
the Polyakov gauge projected U(1) heavy quark potential on them
using time-like Wilson loops. We used the method and computer code
of Heller et al.\ \cite{Heller}. We computed both on axis and off 
axis loops and the effective potential for different time
extensions of the loops have been obtained as
\begin{equation}
   V(R) = - \ln \frac{W(R,T+1)}{W(R,T)}.
\end{equation}
Our results are shown in Figure \ref{fig:pot_pg}.
In the SU(2) case we have a good plateau at $T=3$ (this has also 
been confirmed on another ensemble of larger statistics) but in the
U(1) case the potential decreases considerably with increasing 
$T$ even at this point. Therefore in the SU(2) case we present
the $T=3$ effective potential and for the U(1) case we plot 
the effective potential with several $T$ values. One can
conclude that in the $T \rightarrow \infty$ limit the U(1)
string tension is probably very close to zero. 

The discrepancy is striking. We would also
like to note that the static quark potential measured by Polyakov-loop
correlators is by the very definition of the procedure exactly the
same as the full non-abelian potential. We also note that the string 
tension obtained from Polyakov loop correlators and timelike Wilson loops
should be the same (up to some small finite size effects).
This means that two different  but  physically
equivalent measurements of the same physical quantity give
absolutely different results on the Polyakov gauge projected 
configurations.

For a real test of abelian dominance one has to fix the gauge, do
the abelian projection and show that the resulting abelian 
configurations reproduce all the important long-distance properties
of the non-abelian model. For this test to be nontrivial one
has to include in the measurement a set of operators large enough
so that not all of them can be diagonalised at the same time on
any non-abelian configuration. 

Another important issue about gauge fixing is space-time 
symmetries. Since we want to choose a particular gauge
and not change it when measuring different physical 
quantities and/or under different physical conditions, it seems reasonable
to require that the gauge fixing respect all the space-time  symmetries
of the lattice. Otherwise the abelian projected configurations
would break this symmetry and the resulting continuum theory
would not be euclidean symmetric unless the symmetry is restored 
in some miraculous way. This is also strongly suggested by our
results concerning the Polyakov gauge which treats the time and 
space directions differently. Recently Del Debbio et al.
\cite{Greensite} also showed that asymmetries in the gauge fixing
will lead to similar asymmetries in the abelian projected configurations.
They found that if the maximal abelian gauge fixing is done 
only in a certain plane then abelian dominance holds only for
Wilson loops in that plane.

Some people seem to be troubled by the fact that abelian 
dominance depends on the gauge choice and there are still efforts 
in the literature to prove the contrary \cite{Ejiri}. 
Our result for the Polyakov gauge strongly suggests that
the physics of the abelian projection is not only very strongly
gauge dependent but in most of the arbitrarily chosen gauges 
the abelian projected configurations do not even have a consistent 
physical meaning. In our 
opinion it is only natural to expect that abelian observables 
depend strongly on the gauge fixing and rather than trying to 
find some gauge independence (even in a limited sense) one should
concentrate on finding {\it a particular} gauge in which the
abelian degrees of freedom reproduce as much of the non-abelian 
dynamics as possible.  It is then crucial that the
information discarded in the projection be minimised. This is
exactly what can be achieved with a suitable gauge fixing. The only
well tested method for this is fixing to the maximal abelian 
gauge (MAG) which minimises the off-diagonal components
of the link degrees of freedom, the ones that
are discarded in the projection \cite{MAG}.
This is done by maximising the following quantity:
\begin{equation}
 G[U] = \sum_l \mbox{tr}(U^\dagger_l \sigma_3 U_l \sigma_3),
     \label{eq:GU}
\end{equation}
where the sum runs over all the links, $\sigma_3$,
is a Pauli matrix and $U_l$ is the link SU(2) matrix on $l$. Geometrically
$U^\dagger \sigma_3 U$ can be visualised as a unit vector in the
three-dimensional space spanned by the Pauli matrices. This vector 
is obtained from $\sigma_3$ by applying to it the orthogonal transformation
corresponding to $U$ in the adjoint representation of SU(2). Consequently
the trace which is summed in eq.\ (\ref{eq:GU}) is the projection
of the rotated $\sigma_3$ onto the $\sigma_3$ direction. This quantity 
is maximal when the rotation happens around the $\sigma_3$ axis, i.e.\
$U$ is of the form $\exp(i \alpha \sigma_3/2)$. Maximising $G[U]$ thus 
results in putting all the link matrices as close as possible to this 
diagonal form.  

There might be other gauge choices that preserve the long distance
features better than the MAG but the MAG is the one that -- at least 
locally -- puts as much of the fluctuations as possible into 
the abelian diagonal part of the link variables. For this reason 
the MAG is a priori a better choice than the gauges that diagonalise
an arbitrarily selected set of operators like e.g.\ the Polyakov loops.

\section{Abelian dominance and short-range fluctuations}

In this section we study how abelian dominance in the
maximal abelian gauge depends on the precise nature of 
short distance fluctuations. 

\begin{figure}[!htb]
\begin{center}
\vskip 10mm
\leavevmode
\epsfxsize=110mm
\epsfbox{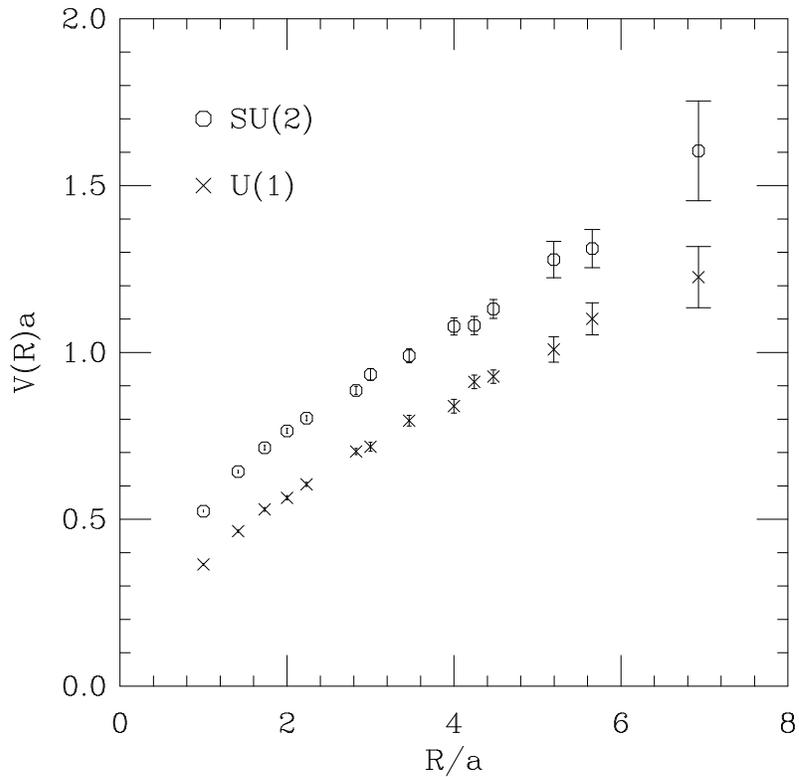}
\end{center}
\caption{The heavy quark potential measured before (SU(2), octagons)
and after (U(1) crosses) abelian projection in the maximal abelian 
gauge on the original configurations.}
\label{fig:pot_mag_b0}
\end{figure}

We generated 100 $8^3 \times 12$ lattices with the fixed 
point action of Ref.\ \cite{DeGrand} at $\beta=1.5$ (lattice 
spacing $a=0.144$ fm). At first as a check we verified that 
abelian dominance holds for this ensemble. We transformed 
the configurations into the maximal abelian gauge maximising 
(\ref{eq:GU}). This was done using the usual overrelaxation 
procedure iterated until the change in the gauge fixing action
became less than $10^{-8}$ per link. After abelian projecting these 
configurations the heavy quark potential was extracted from 
time-like Wilson loops in the same way as in the previous section.
In Fig.\ \ref{fig:pot_mag_b0} this abelian potential is compared 
with the SU(2) potential measured on the same ensemble without
projection. 

A fit to the form 
\begin{equation}
 V(r) = V_0 - \frac{e}{r} + \sigma r
     \label{eq:pot}
\end{equation}
gives $\sigma_{na}=0.123(7)$ for the non-abelian and 
$\sigma_{ab}=0.119(5)$ for the abelian string tension in lattice 
units.

\begin{figure}[!htb]
\begin{center}
\vskip 10mm
\leavevmode
\epsfxsize=110mm
\epsfbox{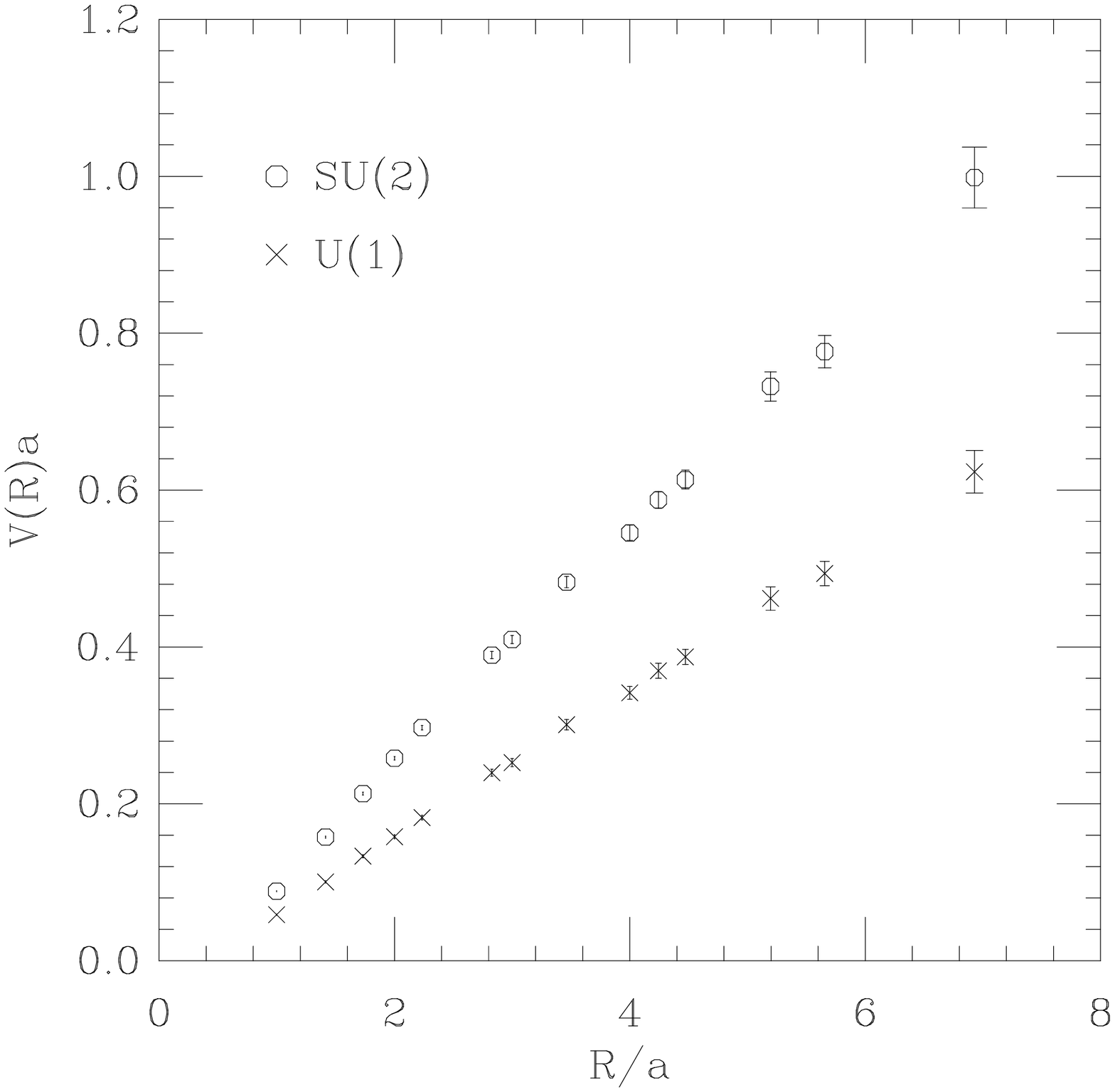}
\end{center}
\caption{The heavy quark potential measured on once smoothed 
configurations. Octagons correspond to the SU(2) potential, 
crosses to the U(1) potential obtained by gauge fixing and 
projecting the smoothed configurations in the maximal 
abelian gauge.}
\label{fig:pot_mag_b1}
\end{figure}

After this check we applied one step of smoothing to the same 
ensemble of SU(2) configuration and repeated the measurement of
the abelian and non-abelian potential on the smoothed 
configurations. The results obtained are shown in Fig.\ 
\ref{fig:pot_mag_b1} and a fit to eq.\ (\ref{eq:pot}) gives
$\sigma_{na}=0.115(9)$ and $\sigma_{ab}=0.080(10)$ for the
SU(2) and the U(1) string tension respectively.

The SU(2) string tension on the smoothed configurations is essentially
the same as on the unsmoothed ones, reflecting the fact that smoothing
does not change the long-distance features. On the other hand, 
as a result of smoothing, the abelian string tension dropped by 
about 30\%. This shows that the abelian string tension is very 
sensitive to the details of the short-distance fluctuations on 
the SU(2) configurations. 

One smoothing step reduces the fluctuations on the length scale of 
the lattice spacing. If several smoothing steps are applied successively,
larger scale fluctuations are also expected to be gradually washed away. 
To check how this affects the abelian string tension, we did another three steps
of smoothing on the SU(2) configurations and after each step both the
abelian and the non-abelian string tension were measured. We found that
neither the abelian nor the non-abelian string tension was changed
by the additional smoothing steps. The stability of the SU(2) string 
tension with respect to smoothing was expected but it is rather surprising
that while the U(1) string tension changed dramatically in the first
step of smoothing, it remained stable after further smoothing.
This suggests that the abelian string tension is a delicate combination 
of the shortest (order $a$) and longest range fluctuations but it is
rather insensitive to intermediate length scales. It seems to us  
quite impossible to reconcile this fact with the expectation that the abelian 
string tension is a genuine long-distance physical observable 
which is in some sense equivalent to the SU(2) string tension. 
In view of this, the approximate abelian dominance found with Wilson 
action in the maximal abelian gauge seems to be an accident rather than
a fundamental physical phenomenon.

\section{Conclusions}

We used a renormalisation group based smoothing to address two questions
related to abelian dominance. Smoothing drastically reduces short distance
fluctuations but it preserves the long distance physical properties of the
SU(2) configurations. This enabled us to extract the
abelian heavy-quark potential from time-like Wilson loops 
on Polyakov gauge projected  configurations. 
We obtained a very small string tension (probably zero).
This is inconsistent with the string tension extracted from Polyakov
loop correlators which trivially reproduces the full SU(2) string tension.
We then argued on general grounds that the only promising gauge
choice in which the abelian projected configurations might capture 
most of the non-abelian physics, is the maximal abelian gauge.

We also applied the smoothing to test how sensitive abelian dominance
in the maximal abelian gauge is to the short distance fluctuations. 
We found that on smoothed SU(2) configurations the abelian string 
tension was about 30\% smaller than the SU(2) string tension which was
unaffected by smoothing. In other words, two ensembles of SU(2)
configurations, having the same long distance physical content
(SU(2) string tension) differring only in the small scale fluctuations,
give different U(1) string tensions. This shows that the abelian 
string tension is not a genuine long distance observable,
it is also very sensitive to the shortest distance scale. If abelian 
dominance is to be regarded as a fundamental phenomenon, it would
be essential to show that it persists in the continuum limit and 
in this limit it becomes independent of the short distance details
of the configurations. Our result suggests that this is quite unlikely
to happen.

In the present paper we did not address the role that 
abelian monopoles might play in the confinement mechanism. In the
recent literature there is a lot of work along this line \cite{mon}
but we feel that the first question one has to ask is whether there is a gauge 
in which the abelian projection reproduces the essential physical 
properties of the non-abelian configurations in a consistent way.
We think that this question has not been unambiguously answered yet.
Until a positive answer to this question is found, 
abelian monopole condensation cannot be accepted as a serious
candidate for explaining confinement.

\section*{Acknowledgements}

T.K. would like to thank Anna Hasenfratz and Tom DeGrand for 
very stimulating discussions and for a careful reading of the 
manuscript. He also thanks Terry Tomboulis and Matt Wingate
for useful conversations and Grigorios Poulis for correspondence. 
We thank the Colorado Experimental High Energy 
Group and the UCLA Elementary Particle Theory group for 
granting us computer time. This work was partially supported by
the Physics Research Group of the Hungarian Academy of Sciences,
Debrecen.
 
\newcommand{\PL}[3]{{Phys. Lett.} {\bf #1} {(19#2)} #3}
\newcommand{\PR}[3]{{Phys. Rev.} {\bf #1} {(19#2)}  #3}
\newcommand{\NP}[3]{{Nucl. Phys.} {\bf #1} {(19#2)} #3}
\newcommand{\PRL}[3]{{Phys. Rev. Lett.} {\bf #1} {(19#2)} #3}
\newcommand{\PREPC}[3]{{Phys. Rep.} {\bf #1} {(19#2)}  #3}
\newcommand{\ZPHYS}[3]{{Z. Phys.} {\bf #1} {(19#2)} #3}
\newcommand{\ANN}[3]{{Ann. Phys. (N.Y.)} {\bf #1} {(19#2)} #3}
\newcommand{\HELV}[3]{{Helv. Phys. Acta} {\bf #1} {(19#2)} #3}
\newcommand{\NC}[3]{{Nuovo Cim.} {\bf #1} {(19#2)} #3}
\newcommand{\CMP}[3]{{Comm. Math. Phys.} {\bf #1} {(19#2)} #3}
\newcommand{\REVMP}[3]{{Rev. Mod. Phys.} {\bf #1} {(19#2)} #3}
\newcommand{\ADD}[3]{{\hspace{.1truecm}}{\bf #1} {(19#2)} #3}
\newcommand{\PA}[3] {{Physica} {\bf #1} {(19#2)} #3}
\newcommand{\JE}[3] {{JETP} {\bf #1} {(19#2)} #3}
\newcommand{\FS}[3] {{Nucl. Phys.} {\bf #1}{[FS#2]} {(19#2)} #3}

\end{document}